\documentclass[sigconf,screen]{acmart} 
\usepackage[capitalise]{cleveref}
\usepackage{multirow}
\usepackage{algorithm}
\usepackage{algpseudocode}
\usepackage{tikz}

\AtBeginDocument{%
  }

\setcopyright{none}
\renewcommand\footnotetextcopyrightpermission[1]{}
\copyrightyear{2024} 
\acmYear{2024} 
\setcopyright{rightsretained} 
\acmConference[ICCAD '24]{IEEE/ACM International Conference on
Computer-Aided Design}{October 27--31, 2024}{New York, NY, USA}
\acmBooktitle{IEEE/ACM International Conference on Computer-Aided Design
(ICCAD '24), October 27--31, 2024, New York, NY, USA}
\acmDOI{10.1145/3676536.3676689}
\acmISBN{979-8-4007-1077-3/24/10}

\begin{document}

\title{Balancing Thermal Relaxation Deviations of Near-Future Quantum Computing Results via Bit-Inverted Programs}


\author{Enhyeok Jang}
\affiliation{
  \institution{Yonsei University}
  \city{Seoul}
  \country{Korea}
}
\email{enhyeok.jang@yonsei.ac.kr}
\orcid{0009-0000-7034-6793}

\author{Youngmin Kim}
\affiliation{
  \institution{Yonsei University}
  \city{Seoul}
  \country{Korea}
}
\email{youngmin.kim@yonsei.ac.kr}
\orcid{0009-0002-8346-4830}

\author{Jeewoo Seo}
\affiliation{
  \institution{Yonsei University}
  \city{Seoul}
  \country{Korea}
}
\email{jiuseo665@gmail.com}
\orcid{0009-0009-8230-158X}

\author{Seungwoo Choi}
\affiliation{
  \institution{Yonsei University}
  \city{Seoul}
  \country{Korea}
}
\email{seungwoo.choi@yonsei.ac.kr}
\orcid{0009-0005-2162-8993}

\author{Won Woo Ro}
\affiliation{
  \institution{Yonsei University}
  \city{Seoul}
  \country{Korea}
}
\email{wro@yonsei.ac.kr}
\orcid{0000-0001-5390-6445}

\renewcommand{\shortauthors}{Enhyeok Jang et al.}

\begin{abstract}

One of the predominant causes of program distortion in the real quantum computing system may be attributed to the probability deviation caused by thermal relaxation. 
We introduce Barber (Balancing reAdout Results using Bit-invErted ciRcuits), a method designed to counteract the asymmetric thermal relaxation deviation and improve the reliability of near-term quantum programs. 
Barber collaborates with a bit-inverted quantum circuit, where the excited quantum state of qubits is assigned to the $\lvert 0 \rangle$ and the unexcited state to the $\lvert 1 \rangle$. 
In doing so, bit-inverted quantum circuits can experience thermal relaxation in the opposite direction compared to standard quantum circuits. 
Barber can effectively suppress the thermal relaxation deviation in program's readout results by selectively merging distributions from the standard and bit-inverted circuits.

\end{abstract}




\maketitle

\section{Introduction}

\noindent 
Quantum computing is anticipated to exceed the capabilities of classical computing across various domains thanks to its ability to exploit unique quantum properties such as superposition and entanglement \cite{khan2024beyond, chae2024elementary, basu2019nist, kole2019improved}. 
However, this potential can be hindered by the noise issue in quantum devices \cite{wang2022quantumnas, park2022fast, oliveiray2023systematic, schmid2024computational}. 
A fundamental strategy for addressing quantum noises involves employing quantum error correction (QEC) \cite{campbell2024series, cirq, bluvstein2024logical}. 
Unfortunately, in near-term quantum devices, the exclusive reliance on QEC may be insufficient for enhancing the fidelity of quantum programs, given the limited availability of qubit resources and the high error rates \cite{bhattacharjee2019muqut}. 
Thus, it is essential to concurrently employ quantum error mitigation techniques with QEC to efficiently suppress errors in qubits \cite{jnane2024quantum, edm}.

One of the primary factors distorting the output distribution in quantum programs is thermal relaxation, which shortens the coherence time of qubits in current quantum devices \cite{huang2024high, zhao2024dual, zulehner2020approximation}. 
Due to the thermal relaxation, resulting distributions of quantum circuit executions show asymmetric measurement errors where misreading $\lvert1\rangle$ to $\lvert0\rangle$ is more frequent than vice versa \cite{qcuk,tannu3, yayla2021fefet, utt2024quantum}.
According to our experiments with the 12-qubit GHZ (Greenberger-Horne-Zeilinger \cite{ghz, peng2023short}) state circuit on the latest 127-qubit \textit{ibm\_nazca} quantum systems \cite{ibm}, the all-zero state can be measured up to 67\% more frequently than the all-one state.
Reducing the probability distortions between correct answer states is pivotal for correctly interpreting the outcomes of quantum computing.
This is because in many quantum algorithms such as HHL (Harrow-Hassidim-Lloyd) \cite{harrow2009quantum} and VQE (Variational Quantum Eigensolver) \cite{vqe}, probability ratios between correct answer states become a solution to the problem directly.
Consequently, this study focuses on addressing relaxation deviation to enhance the reliability of quantum programs.

Recently proposed Invert-and-Measure leverages the bias property of this asymmetric measurement error, improving the reliability of quantum programs \cite{tannu3}.
By inverting qubits just before the readout process, Invert-and-Measure turns the vulnerable state into a robust one for the quantum system to output.
Unfortunately, according to experiments using quantum workloads requiring multiple correct answer states, such as GHZ state circuits \cite{ghz}, Grover's search circuit \cite{grover}, or QAOA (Quantum Approximate Optimization Algorithm) for Max-Cut circuit \cite{farhi, jang2024recompiling}, we observe that Invert-and-Measure may amplify the probability deviation between answers.

The reasons for these results can be analyzed as follows: The asymmetric measurement error in quantum programs is mainly due to the thermal relaxation applied in the entire process of quantum circuit execution.
Since the inverted circuit design proposed by Invert-and-Measure reverses qubits just before the readout process, the inverted circuit execution experiences thermal relaxation in almost the same direction as the standard circuit, which may raise concerns that probability deviations between correct answers might be augmented when merging standard and inverted execution results.
Thus, if the inverted circuit is designed to experience thermal relaxation in the opposite direction to the standard circuit, the probability deviation between correct answer states can be efficiently offset by merging the standard and bit-inverted execution results.

To address the probability deviation in quantum program results, we introduce Barber (Balancing reAdout Results using Bit-invErted ciRcuits). 
Barber employs the bit-inverted quantum circuit, where the excited state of qubits is designated as the $\lvert0\rangle$, and the unexcited state of qubits is designated as the $\lvert1\rangle$ (which is the reverse of the standard configuration). 
This arrangement enables bit-inverted quantum circuits to experience thermal relaxation in the opposite direction of standard quantum circuits.
For example, in the bit-inverted configuration, the probability of $\lvert0\rangle$ being mismeasured as the $\lvert1\rangle$ can be higher than that of the $\lvert1\rangle$ being mismeasured as $\lvert0\rangle$.

It should be noted that standard gate operations cannot be utilized as they are when the defined basis changes.
To implement bit-inverted circuits, Barber employs a distinct gate operation approach. 
Gates for bit-inverted circuits can be realized by appending a Pauli-X gate before and after each standard gate operation.
Barber then synthesizes the final probability distribution by selectively merging and normalizing results from both standard and bit-inverted circuit measurements. 
By doing this, Barber effectively offsets the thermal relaxation error, enhancing the reliability of the quantum program.

The main contributions in this research (Barber) are as follows:
\begin{itemize}
        \item We observe that thermal relaxation errors can distort the output distributions of quantum programs, and existing mitigation techniques may not address this problem sufficiently.
        \item We propose Barber, which leverages bit-inverted circuits that experience the effects of thermal relaxation inversely to offset probability deviations between correct answer states.
        \item Leveraging gate pruning technique, the bit-inverted quantum circuit can be synthesized with less than an average of 5\% additional depth overhead than the standard execution.
        \item By selectively merging the probabilities of standard and bit-inverted programs, our proposed output reconstruction achieves the 2.1$\times$ speed-up compared to Invert-and-Measure.
        \item Barber reduces the probability deviation between correct answer states by 86\%, improves PST by 72\%, and decreases the Hellinger distance by 60\% than the standard executions.
        \item Compared with state-of-the-art quantum error mitigation techniques, Enhancing Virtual Distillation, Invert-and-Mea-sure, and Quixote, Barber achieves the highest PST and lowest Hellinger distance from the ideal probability distribution.
        \item Barber's method can be applied to all kinds of gate-based quantum computers (superconducting \cite{ibm}, neutral atom \cite{jang2025qubit}, etc.) without requiring any modifications of hardware.
\end{itemize}


\section{Motivation}

\subsection{Relaxation Error in Quantum Computing}

\noindent 
In this work, we focus on the thermal relaxation error, which is one of the dominant sources of error in quantum programs \cite{tannu3, wille2020efficient, ibm, qcuk}.
Due to various noise sources and interactions with the surrounding environment, qubits lose their coherence and collapse into $\lvert0\rangle$ state \cite{qcuk}. 
The time constant \textit{t1} shows how quickly a qubit loses a superposed or excited state due to the thermal relaxation \cite{baheri2022pinpointing}.
For example, the equation for the probability of a qubit remaining in the excited state ($P(t)$) at time $t$ is given by $P(t) = e^{-t/t1}$.
\textit{t1} is a critical parameter for determining the error rate of quantum computers \cite{reinhold}.
For instance, \textit{t1} of qubits on the IBM's 65-qubit \textit{Hummingbird r3} processor is approximately 3 to 280 microseconds, and that on the IBM's 127-qubit \textit{Eagle r3} processor is approximately 20 to 500 microseconds \cite{ibm}.
\textit{t1} of the Rigetti's \textit{Aspen} processors are known to be 8 to 33 microseconds \cite{rigetti}, and \textit{t1} of Baidu's \textit{Qian Shi} processor is known to be approximately 30 microseconds \cite{rqa}.

\subsection{Error Impact Analyses by Relaxation}

\begin{figure} [ht] 
\centerline {
\includegraphics [width=\columnwidth] {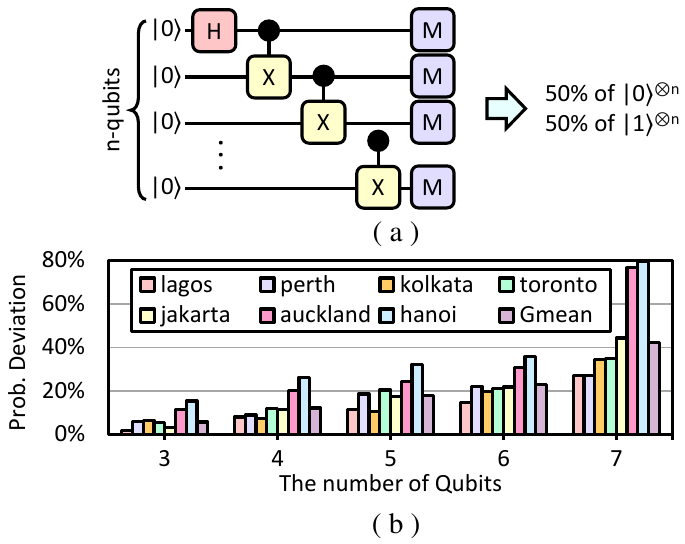} }
\caption {
(a) shows the $n$-qubit GHZ-state circuit and its expected noise-free measurement results (All-zero and all-one states are measured by 50\%, respectively).
(b) shows the probability deviation between all-zero states and all-one states measured by IBM quantum systems.
Experiments were conducted on IBM quantum systems, calibrated on Jan. 5, 2023.
} 
\label{f0} 
\end{figure}

\noindent
Our experimental observation indicates that thermal relaxation can significantly distort the probability distribution of quantum program results, particularly in programs with multiple correct answers.
For example, in the GHZ (Greenberger-Horne-Zeilinger \cite{ghz}) state, the probability of measuring all-zero states is expected to be equal to the probability of measuring all-one states in the absence of noise, as shown in \cref{f0} (a).
However, due to thermal relaxation, the probability of measuring all-zero states becomes higher than all-one states.
The deviation between these two correct answers tends to grow more significant with increasing the number of qubits, and in the worst-case, all-zero states are observed to be measured about 80\% higher than all-one states, in the 7-qubit GHZ state circuit, as shown in the case of \textit{ibm\_hanoi} system in \cref{f0} (b).

Additionally, we analyzed the probability deviation between correct answer states of 12-qubit GHZ state circuits \cite{ghz} by changing qubit groups in \textit{ibm\_nazca} system \cite{ibm}, as shown in \cref{t0}.
Experiments in \cref{t0} show that the probability deviation between correct answer states can be significantly increased depending on the geometric mean of \textit{t1} time by selected qubits.
These experimental results imply that the probability deviations between correct answer states could be significantly affected by thermal relaxation.

\begin{table}[ht]
        \centering
        \renewcommand{\arraystretch}{1}
        \renewcommand{\tabcolsep}{0.6mm}
        \caption{
        Probability Deviations of 12-Q GHZ According to Qubit Groups (Experiments were conducted on May 3, 2024)}
        \label{t0}
        \begin{tabular}{|c|c|c|c|c|}
        \hline
        Qubits in Groups (ibm\_nazca) & \textbf{Gmean of t1} & PST & \textbf{Deviation} \\
        \hline
        4-10, 15, 22-24, 34 & 221 (us) & 0.35 & \textbf{13.4\%} \\
        0-11 & 217 (us) & 0.32 & \textbf{19.3\%} \\
        78-83, 92, 102-106 & 213 (us) & 0.31 & \textbf{21.1\%} \\
        71, 77-79, 91, 96-98, 109, 114-116 & 204 (us) & 0.29 & \textbf{31.0\%} \\
        72, 81-83, 87, 92-93, 102-106 & 181 (us) & 0.20 & \textbf{67.3\%} \\
        \hline
        \end{tabular}
\end{table}

The deviation between correct answer states may lead to incorrect interpretations of quantum program results, particularly in algorithms where the probability differences between correct answer states are crucial for solving the problem.
Note that the probability differences between correct answer states hold significance in numerous renowned quantum algorithms. 
Probability differences between readout results may be the core information in computational results by algorithms.
For instance, in the HHL (Harrow-Hassidim-Lloyd) algorithm, the probability ratio between correct answer states directly offers solutions to linear algebraic operations \cite{babukhin2023harrow, harrow2009quantum}. 
Similarly, in the VQE (Variational Quantum Eigensolver), the probability of an electron existing in each orbital is directly relevant to the results \cite{kim2024variational}.
In conclusion, the probability differences between correct answer states are pivotal for interpreting quantum program outcomes, underscoring the importance of research aimed at minimizing deviations between correct answers.

\subsection{Previous Study for Asymmetric Bit Error}

\noindent 
Representative studies addressing the asymmetric bit error \cite{wen2013cd, amrouch2021brain, li2017improving, gherman2020binary} for quantum programs include Invert-and-Measure \cite{tannu3}.
Invert-and-Measure introduces an inverted circuit (implemented by adding Pauli-X gates to each qubit immediately before readout) to counteract the asymmetric measurement error. 
By merging the results from standard and inverted circuits, Invert-and-Measure effectively improves PST for quantum program results.
However, in some workloads with multiple correct answers, we observed that Invert-and-Measure may exacerbate the deviation between answers when combining results from the standard and inverted programs.

\begin{figure} [ht] 
\centerline {
\includegraphics [width=\columnwidth] {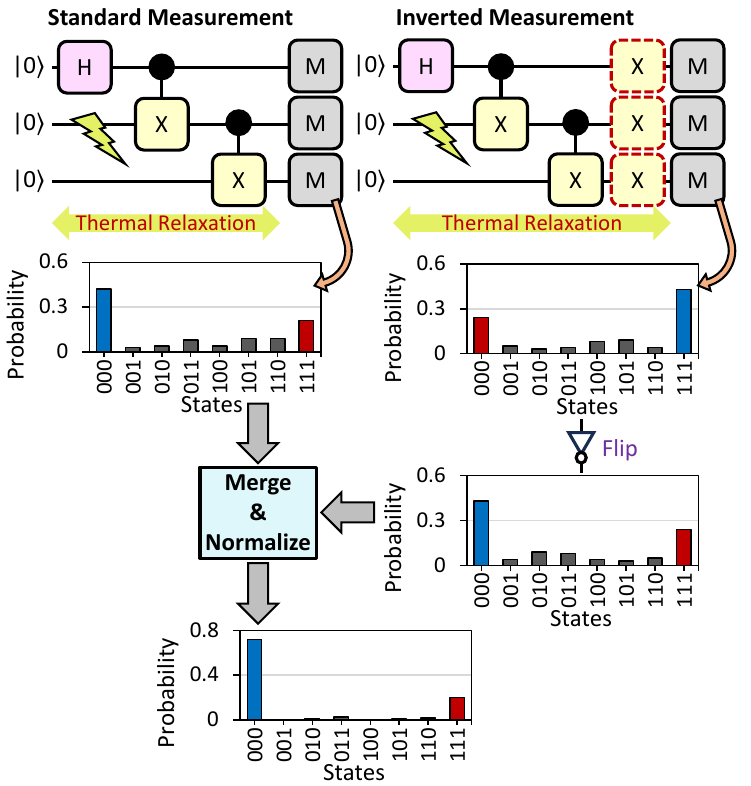} }
\caption {
An overview of applying Invet-and-Measure to the 3-qubit GHZ state circuit. 
Among the two correct answer states, the blue bar refers to the relatively robust state in quantum systems to the thermal relaxation error, and the red bar refers to relatively vulnerable state (answer) in quantum systems.
} 
\label{f6} 
\end{figure}

As an example, \cref{f6} shows a scenario in which deviations between correct answer states may be augmented in 3-qubit GHZ state circuits.
Note that the ideal measurement results for the GHZ state circuit are 50\% for all-zero and all-one states.
In the standard circuit execution, all-zero states can be measured more frequently than all-one states due to thermal relaxation.
The inverted circuit execution can measure all-one states more frequently than all-zero states since the robust state and the vulnerable state are exchanged by the Pauli-X gates applied just before the measurement.
However, note that if each state bit is flipped to correctly interpret the results measured from the inverted circuit, the all-zero state will have a higher probability than the all-one state again.
Consequently, when merging the results from standard and inverted circuits, the probability deviation between correct answer states may be amplified.

\begin{figure} [ht] 
\centerline {
\includegraphics [width=\columnwidth] {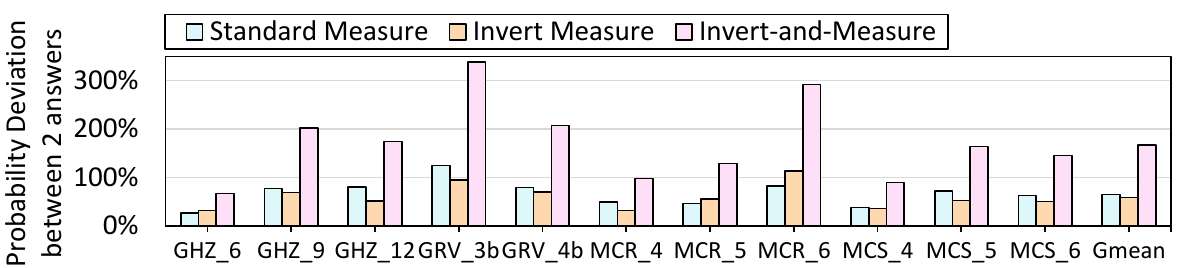} }
\caption {
Comparison of probability deviations between correct answer states of each measurement method according to quantum workloads with two correct answer states with equal probability on the \textit{ibm\_algiers} system (lower is better).
} 
\label{f7} 
\end{figure}

To quantitatively evaluate the deviation between correct answers according to the quantum program execution methods, we conducted experiments on various 2-answer quantum circuits on the \textit{ibm\_algiers} system.
The results are shown in \cref{f7}, where the deviation between correct answer states is compared for standard, inverted, and Inverted-and-Measure methods.
Experimental results in \cref{f7} show that applying Invert-and-Measure can augment the deviation between correct answers of 2-answer workloads with these identical probabilities by an average of 147\% and up to 251\%.

We carefully hypothesize and analyze the phenomenon of augmenting deviations between correct answers as follows:
(i) The errors caused by thermal relaxation occur throughout the execution of quantum circuits.
(ii) In Invert-and-Measure, the inverted circuit experiences thermal relaxation in the same direction as the standard circuit for most of its running time.
(iii) Thus, the process of merging the standard circuit and the inversion circuit may amplify the deviation between the correct answer states.
This means that the inversion technique with Pauli-X gates just before the measurement may not be sufficient to compensate for the thermal relaxation deviations.
Therefore, it is expected that if the inverted circuit is designed to experience thermal relaxation in the opposite direction to the standard circuit in the entire process of executions, the deviations between correct answers can be offset by each other.

\subsection{Lack of Mitigating Relaxation Error Studies}

\noindent
Research topics aimed at mitigating errors in quantum program results have been extensively studied due to their importance \cite{jigsaw, tannu3, maksymov2024utilizing, jang, dangwal2023varsaw}. 
However, to the best of our knowledge, there are few studies that address probability deviations by the thermal relaxation of quantum programs.
We hypothesize three reasons for the scarcity of these studies:
(i) At the time of previous research, the noise in devices was too severe, and errors caused by thermal relaxation may not have been distinctly recognized.
For example, experimental results from Invert-and-Measure show that the IST (Inference Strength \cite{oliveiray2023systematic}) of the 4-qubit Bernstein-Vazirani (BV \cite{bv}) circuit execution result performed on the ibmqx4 processor \cite{ibm} was lower than 0.5 \cite{tannu3}, but the IST of the 8-qubit BV circuit execution result performed on the IBM Falcon r5.11 processor \cite{ibm} we performed is higher than 1.
This implies that in experiments using early processors such as ibmqx4, the error characteristics between correct answer states due to thermal relaxation may have been difficult to notice clearly.
(ii) Quantum circuit benchmarks in previous quantum error mitigation research mainly focused on typically include BV circuit, Grover's circuit for searching a single state \cite{grover}, or reversible logic circuits \cite{revlib, datta2014post, kole2016heuristic, marbaniang2017design, yamashita2010fast}. 
These scenarios have only a single answer, which might lead to insufficient consideration of the balance between correct answer states.
(iii) Standard metrics used to quantitatively evaluate the accuracy of quantum program results (such as PST \cite{vadali2024quantum, khadirsharbiyani} and IST \cite{oliveiray2023systematic}) primarily focus on the ratio between correct and incorrect answers and may not adequately reflect errors arising from deviations between answers.

Consequently, the issue of the probability deviation between correct answer states may have received relatively less attention.



\section{Barber Design Methodology}

\subsection{Design Challenges for Implementing Barber}

\noindent
The four design challenges for Barber methodology are as follows.

(i) The inverted quantum circuit should be designed to experience thermal relaxation in the opposite direction to the standard quantum circuit, thereby efficiently mitigating the probability deviation between correct answers:
To address this challenge, Barber introduces a bit-inverted quantum circuit.
In the bit-inverted circuit, the excited state of qubits is assigned to the $\lvert0\rangle$, and the unexcited state of qubits is assigned to the $\lvert1\rangle$, which is the reverse of the standard qubit configuration.
Therefore, the bit-inverted quantum circuit can experience thermal relaxation in the opposite direction of standard quantum circuits.
By merging the output distributions of the standard and the bit-inverted circuit, Barber can efficiently offset the probability deviation between the correct answer states.

(ii) Note that the bit-inverted quantum circuits discussed in (i) cannot apply standard quantum gates for standard circuits.
In other words, in a bit-inverted circuit, the defined basis is the opposite, requiring a redefinition of the quantum gate:
To address this challenge, Barber introduces \textit{Gate Inversion} technique for bit-inverted circuits.
The bit-inverted circuit can be implemented by attaching $n$ Pauli-X gates to both sides of $n$-qubit gate of the standard circuit.

(iii) It should be noted that an increase in quantum gates can lead to an increase in error rates in near-term quantum devices, resulting in a decrease in output accuracy. 
Therefore, the additional quantum gates required in the implementation of the bit-inverted circuit discussed in (ii) should be minimized:
Fortunately, the additional Pauli-X gates in the bit-inverted circuit discussed in (ii) can be canceled with other Pauli-Xs added to the adjacent gate, which is possible because the Pauli-X operator is a Hermitian matrix \cite{chuang}.
By introducing the Gate Pruning technique, Barber efficiently minimizes the added Pauli-X gates except for the firstly applicated ones for initialization, which reverses the thermal relaxation direction.

(iv) The probability distribution reconstruction process should be efficient and fast:
To address this challenge, Barber introduces \textit{Selective Merge \& Normalize} technique, which is based on the observation that if the size of the probabilities is zero or very small, skipping them can contribute to reducing reconstruction time for output distribution.
Barber merges the outputs of the standard and bit-inverted circuits whose probability is larger than a defined threshold and normalizes the probabilities of the selected states by the sum of their probabilities.
By doing so, Barber can efficiently reconstruct the probability distribution with little loss of accuracy.

\subsection{Barber Overview} 

\begin{figure} [h] 
\centerline {
\includegraphics [width=\columnwidth] {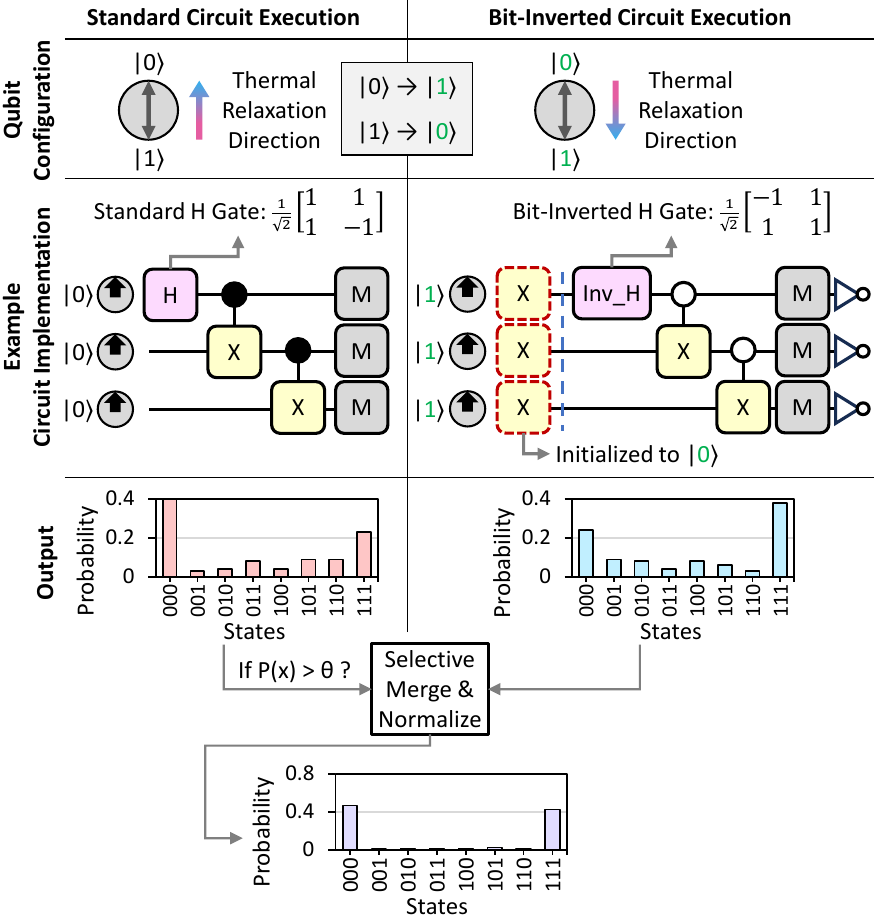} }
\caption {
An overview of Barber using a 3Q GHZ state circuit.
} 
\label{f1} 
\end{figure}

\noindent
\cref{f1} shows an example of Barber execution on a 3-qubit GHZ state circuit.
Note that the ideal noise-free results of this GHZ circuit should be measured by $\lvert 000 \rangle$ and $\lvert 111 \rangle$ at 50\% probability, respectively.
Unfortunately, thermal relaxation errors can cause $\lvert 111 \rangle$ to be measured with lower probability, as shown in the standard circuit execution in \cref{f1}.
To address this challenge, Barber exploits the bit-inverted quantum circuit.
For bit-inverted circuits, the logical basis of qubits is defined in the opposite direction, which makes the bit-inverted circuit can experience thermal relaxation in the opposite direction.
In other words, one of the correct answers that is measured as having a lower probability in standard circuit execution can be measured relatively higher in bit-inverted circuits and vice versa.
By merging the output distributions of the standard and bit-inverted circuits, Barber effectively offsets the probability deviation between the correct answer states, as shown in the \cref{f1}.

The unitary operation for bit-inverted gates can be implemented by taking the Pauli-X operator on both sides of the unitary from its standard version.
Barber leverages Gate Pruning technique to minimize the additional depth of bit-inverted circuits.
Our evaluation shows that the depth of the bit-inverted circuit does not exceed 5\% in the geometric mean than the standard circuit.
This is because the extra Pauli-X arising from the bit-inverted circuit can be canceled efficiently since the Pauli-X operator is an Hermitian \cite{chuang}.
Finally, Barber employs \textit{Selectively Merge \& Normalize}, which merges the outputs of the standard and bit-inverted circuits whose probability is larger than a defined threshold and normalizes the probabilities of the selected states by the sum of their probabilities.
This reconstruction method achieves 2.1$\times$ speed-up than the \textit{Merge \& Normalize} by Invert-and-Measure, with a little loss of accuracy.

\subsection{Bit-Inverted Qubit Configuration} \label{biqc}

\noindent
This section describes the qubit configuration of the bit-inverted quantum circuit employed in Barber.
Note that the logical basis of qubits in the bit-inverted circuit is defined in the opposite direction from the standard quantum circuit.
In other words, the unexcited or ground state is defined as $\lvert1\rangle$ in the bit-inverted quantum circuit.
Hence, the order of the bit-inverted state amplitudes describing the state vector is contrary to that of the standard state amplitudes.
For example, assuming an $n$-qubit system, the standard state amplitude order and the bit-inverted state amplitude order are as follows.
\begin{equation*}
\centering
\lvert \psi_{standard} \rangle = 
\begin{bmatrix}
        a_0 \\
        a_1 \\
        ... \\
        a_{2^{n-1}}
\end{bmatrix}, \quad
\lvert \psi_{Bit-Inverted} \rangle = 
\begin{bmatrix}
        a_{2^{n-1}} \\
        a_{2^{n-2}} \\
        ... \\
        a_0
\end{bmatrix}.
\end{equation*}

Quantum programs typically start at all zero states.
For circuit execution on the bit-inverted system, all qubit states can be initialized to zero by applying Pauli-X gates to all qubits.
By doing so, the bit-inverted circuit could experience the thermal relaxation error in the opposite direction to the standard quantum circuit execution.

\subsection{Unitary Preparation for Inverted Programs} \label{inv}

\noindent
In this section, we explain the Gate Inversion process for implementing quantum gates using the bit-inverted quantum circuit.
As discussed in \cref{biqc}, the logical basis of qubits in the bit-inverted quantum circuit is the opposite of standard quantum circuits since the ground state is defined as $\lvert1\rangle$.
The elements of the unitary matrix for quantum gates for use on the bit-inverted system should be modified since the order of state amplitudes is defined differently.
In other words, quantum gates for bit-inverted circuits should be transformed to fit the bit-inverted basis. 
The unitary operator corresponding to these bit-inverted quantum gates could be obtained by applying the Pauli-X operator \cite{chuang} to both sides.
For instance, assuming a unitary matrix U$_G$ describing an $n$-qubit gate G, the gate G should be modified to the unitary matrix U$_{inv\_G}$ from the gate inv\_G (inverted G) for applying to the bit-inverted circuit as:
\begin{equation*} 
\centering
U_{Inv\_G} = X^{\otimes n} U_{G} X^{\otimes n}.
\end{equation*}

For example, as also described in \cref{f1}, the unitary matrices for Pauli-X and Hadamard gates \cite{chuang} could be modified for applying to bit-inverted quantum program execution as follows, respectively.
\begin{equation*}
\centering
U_{Inv\_X} = XXX =  
\begin{bmatrix}
        0 & 1 \\
        1 & 0 \\
\end{bmatrix}, \quad
U_{Inv\_H} = XHX = \cfrac{1}{\sqrt{2}}
\begin{bmatrix}
        -1 & 1 \\
        1 & 1 \\
\end{bmatrix}.
\end{equation*}

As shown above, the unit matrix of the Pauli-X gate for using the standard and inverted basis is identical.
This means that the Pauli-X gate can be applied to both standard and bit-inverted circuits without any modification.
On the other hand, the unitary matrix for the Hadamard gate to be leveraged in bit-inverted circuits should be modified.
In terms of quantum circuit execution, if quantum hardware does not support these modified operations natively, it should be implemented by attaching an X gate to both sides of the existing standard gate.
For another example, in the case of CNOT, the unitary should be modified for use in bit-inverted circuits as:
\begin{equation*}
\centering
U_{Inv\_CNOT} = (X \otimes X) CNOT (X \otimes X)
=
\begin{bmatrix}
        0 & 1 & 0 & 0 \\
        1 & 0 & 0 & 0 \\
        0 & 0 & 1 & 0 \\
        0 & 0 & 0 & 1 \\
\end{bmatrix}.
\end{equation*}

The corresponding quantum gate to this implementation is a zero-controlled-NOT gate (or a negatively controlled-NOT \cite{chuang}).

\subsection{Pruning and Barrier for Inverted Initializing}

\noindent
As discussed in \cref{inv}, if the bit-inverted gate is not supported on quantum hardware natively, the Gate Inversion process requires adding Pauli-X gates to both sides of each gate from the standard circuit.
There can be a risk of increasing gate errors and reducing fidelity running in quantum devices if the circuits with these added Pauli-X gates are executed as they are.
Fortunately, two consecutive Pauli-X gates can be offset since Pauli-X is a Hermitian matrix, and they become an identity operation.
The gate cancellation technique can be supported by most quantum computing compilers.
For example, in the case of the IBM Qiskit Transpiler, this technique could be supported by \textit{Communicative Cancellation Initializer} function.

It should be noted that when performing the Gate Pruning, Pauli-X gates applied to initialize to zero for the bit-inverted qubits at the beginning of the circuit should not be canceled since they contribute to experiencing thermal relaxation in the opposite direction.
This problem can be addressed by applying the \textit{Barrier} instruction after the first applied Pauli-X gates.
By doing so, Pauli-Xs for initialization of bit-inverted qubits are not canceled by Gate Pruning.

\begin{figure} [h] 
\centerline {
\includegraphics [width=\columnwidth] {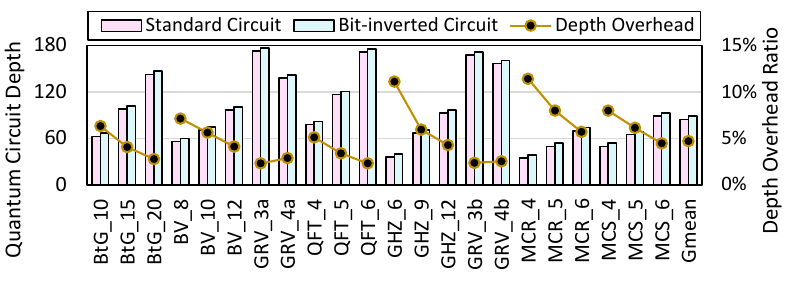} }
\caption {
Circuit depth comparisons and their depth overhead analysis of the standard and proposed bit-inverted circuits.
} 
\label{f8} 
\end{figure}

\cref{f8} shows the circuit depth comparison and overhead analysis of the standard and proposed bit-inverted quantum circuits (Gate Pruning is applied.) on the IBM Eagle r3 processor using various quantum benchmarks \cite{ibm}.
Our evaluation shows that the depth-increasing overhead of bit-inverted circuits increases by a geometric mean of 4.7\% and up to 11.4\% compared to standard circuit execution.
As the Pauli-X gates required to implement the bit-inverted circuit cancel each other efficiently, only constant depth overhead is required regardless of the scale of the quantum program.
Thus, the additional depth overhead ratio is observed to become smaller as the circuit depth increases.
Thus, the execution cost of the bit-inverted quantum circuit does not remarkably increase when compared to the standard circuit and that these additional depth overheads become negligibly small if the quantum program sufficiently grows in depth.
These results implies that the bit-inverted circuit can be executed without a significant increase in execution cost, and the Gate Pruning can efficiently minimize the additional depth overhead.

\subsection{Selective Merging and Normalizing Process}

\noindent
Barber operates by executing standard and bit-inverted quantum circuits using half the original number of measurement shots for each, which follows Invert-and-Measure's shot assignment policy \cite{tannu3}.
To correctly interpret the results from the bit-inverted circuit, a bit-flip is applied to each output. 
Barber then selectively combines and normalizes states whose probabilities exceed a threshold ($\theta$) set for the outcomes of the standard circuit. 
The threshold $\theta$ is assumed as $1/N^2$, where $N$ represents the size of the quantum state space. 
\textit{Selective Merge \& Normalize} skips the reconstruction process for probabilities that are negligibly small, thereby streamlining the output reconstruction time needed to generate the result distribution. 
Probabilities below the threshold are obtained by utilizing the probability results of standard circuit execution.
The sum of the probabilities of the total resulting distribution is conserved by 1 since the probabilities of the selected states are normalized by the sum of their probabilities.
Consequently, the final probability distribution is reconstructed by selectively merging the outputs from the standard and bit-inverted quantum circuits, thereby effectively offsetting the impact of the thermal relaxation error in programs.

\subsection{Output Reconstruction Overhead Analysis}

\begin{figure} [h] 
\centerline {
\includegraphics [width=\columnwidth] {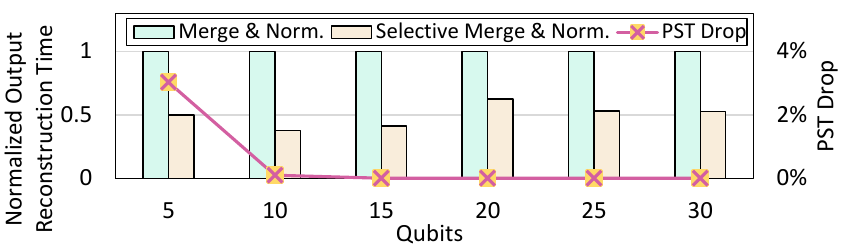} }
\caption {
Comparison of normalized probability distribution generation time according to the reconstruction method: \textit{Merge \& Normalize} from Invert-and-Measure and \textit{Selective Merge \& Normalize} from the Barber method (lower is better).
} 
\label{f9} 
\end{figure}

\noindent
This section compares the performance of \textit{Selective Merge \& Normalize} by Barber over \textit{Merge \& Normalize} by Invert-and-Measure \cite{tannu3}.
\cref{f9} shows the comparison of the normalized output reconstruction time according to these two methods and the PST drop ratio by \textit{Selective Merge \& Normalize} over \textit{Merge \& Normalize}.
The experiment in \cref{f9} was performed using a single CPU (Intel Core i9-9900K) environment with 64 GB memory.
For a fair comparison of the PST drop, both reconstruction methods employ identical probability distributions of standard and bit-inverted circuits performed by Barber; thus, there is only a difference in the reconstruction of output distributions.
The probability distributions were obtained by $n$-qubit GHZ state circuits on \textit{ibm\_kyoto} system with 100,000 shots \cite{ibm}.
According to our evaluation, \textit{Selective Merge \& Normalize} achieves a speedup of 2.1$\times$ in the geometric mean compared to \textit{Merge \&Normalize} in the range of 5 to 30 qubits.
The PST drop decreases continuously as the qubit number increases, and at 20 qubits, it drops to about $3\times10^{-5}$.
These experimental results demonstrate that Barber's probability distribution reconstruction method improves the speed by more than a factor of 2 with little loss of accuracy compared to the previously proposed probability distribution reconstruction method by the Invert-and-Measure.

\section{Experimental Methodology}

\subsection{Quantum Program Execution Scenario}

\noindent 
Barber is compared with various error mitigation techniques.
We leverage Qiskit Runtime Sampler V1 with default resilience setting and optimization level 3 \cite{ibm}.
Following scenarios are considered.

\textbf{Standard}: this refers to the standard circuit execution version.

\textbf{Enhancing Virtual Distillation} (EVD) \cite{li2023enhancing}: This refers to the quantum circuit execution version that applies the virtual distillation technique \cite{huggins2021virtual} with the quantum circuit-cutting methodology.

\textbf{Invert-and-Measure} \cite{tannu3}: refers the execution version that applies the Invert-and-Measure technique for the quantum error mitigation.
Adaptive Invert-and-Measure (AIM) scheme is applied.

\textbf{Quixote} (Quantum Independent Execution Architecture) \cite{jang}: This refers to a quantum circuit execution version that applies the Quixote's circuit partitioning technique for error mitigation.
The number of sub-circuits considered is 2 in common for all workloads.

\textbf{Barber}: refers to a version of our technique for error mitigation.

\subsection{Quantum Backend and Benchmark Circuits}

Various benchmarks were evaluated on the IBM quantum system.

\textbf{Quantum Backend}:
The experiments were conducted on April 29, 2024, utilizing the 27-qubit \textit{ibm\_algiers} system calibrated at that time \cite{ibm}.
The number of measurement shots for each quantum circuit execution to obtain the output distributions is 1,024 times.

\textbf{Benchmark Quantum Circuits}:
Various benchmark quantum circuits used in these experimental evaluations are shown in \cref{t1}, where the number of correct answer states includes one and two.

\textbf{Setup for QAOA Circuits}: 
We leverage the classical quantum circuit simulator (Pennylane \texttt{default\_qubit} \cite{bergholm}) for utilizing trained QAOA (Quantum Approximate Optimization Algorithm \cite{farhi}) circuits.
The parameters in QAOA circuits were updated using PennyLane Optimizer \cite{bergholm}, and the parameter variables used in experiments are presented in \cref{t2}.
The DFS (Depth First Search)-based method \cite{majumdar} and the Intelligent Re-order \cite{alam2020circuit} were applied.

\begin{table}[t]
\centering
\renewcommand{\arraystretch}{1}
\renewcommand{\tabcolsep}{0.5mm}
\caption{
Evaluated benchmark Quantum circuits}
\label{t1}
\begin{tabular}{|c|c|c|c|c|c|}
\hline
\multirow{2}{*}{Workload} & Algorithm & \multirow{2}{*}{Qubits} & Correct & 1Q & Multi-Q  \\
 & [Reference] & & Answer(s) & Gates & Gates  \\
\hline
\hline
BtG\_10 & Binary & 10 & 0x2aa & 5 & 9 \\
BtG\_15 & to Gray code & 15 & 0x0aaa & 7 & 14 \\
BtG\_20  & conversion \cite{revlib} & 20 & 0xaaaaa & 10 & 19 \\
\hline
BV\_8 & Bernstein- & 8 & 0xff & 17 & 7 \\
BV\_10 & Vazirani & 10 & 0x3ff & 21 & 9 \\
BV\_12  & algorithm \cite{bv} & 12 & 0xfff & 25 & 11 \\
\hline
GRV\_3a & Grover's & 3 & 0x7 & 39 & 4 \\
GRV\_4a  & algorithm \cite{grover} & 4 & 0xf & 76 & 6 \\
\hline
QFT\_4 & Quantum & 4 & 0xf & 12 & 8 \\
QFT\_5 & Fourier & 5 & 0x1f & 15 & 12 \\
QFT\_6  & Transform \cite{shor} & 6 & 0x3f & 18 & 18 \\
\hline
\hline
GHZ\_6 & Greenberger- & 6 & 0x00, 0x3f & 1 & 5 \\
GHZ\_9 & Horne–Zeilinger & 9 & 0x000, 0x1ff & 1 & 8 \\
GHZ\_12  & state \cite{ghz} & 12 & 0x000, 0xfff & 1 & 11 \\
\hline
GRV\_3b & Grover's & 3 & 0x0, 0x7 & 21 & 3 \\
GRV\_4b  & algorithm \cite{grover} & 4 & 0x0, 0xf & 52 & 6 \\
\hline
MCR\_4 & QAOA for & 4 & 0x5, 0xa & 12 & 8 \\
MCR\_5 & 1-layer Max-Cut & 5 & 0x0a, 0x15 & 15 & 10 \\
MCR\_6  & (ring graph) \cite{farhi} & 6 & 0x05, 0x2a & 18 & 12 \\
\hline
MCS\_4 & QAOA for & 4 & 0x7, 0x8 & 11 & 6 \\
MCS\_5 & 1-layer Max-Cut & 5 & 0x0f, 0x10 & 14 & 8 \\
MCS\_6  & (star graph) \cite{farhi} & 6 & 0x1f, 0x20 & 17 & 10 \\
\hline
\end{tabular}
\end{table}

\begin{table}[h]
\centering
\renewcommand{\arraystretch}{1}
\renewcommand{\tabcolsep}{0.7mm}
\caption{
Simulated Parameters for Evaluated QAOA Circuits}
\label{t2}
\begin{tabular}{|c|c|c|c|c|c|c|}
\hline
Parameters & MCR\_4 & MCR\_5 & MCR\_6 & MCS\_4 & MCS\_5 & MCS\_6 \\
\hline
\hline
$\lambda$ (rad) & -0.8 & -0.78 & -0.79 & -0.93 & -1.1 & -1.57 \\
$2\beta$ (rad)  & 0.79 & 0.78 & 0.78 & 0.78 & 0.79 & 0.79 \\
\hline
\end{tabular}
\end{table}

\subsection{Evaluation Metrics}

3 metrics are utilized for quantitative evaluations of program results.

\textbf{Probability Deviation}: This metric refers to the probability deviations between correct states and is used for evaluating quantum benchmark circuits with two correct answers.
Probability Deviation is given as $(a-b)/b \times 100\%$, where $a$ and $b$ are the probabilities of the two correct answer states with $a \geq b$.
Thus, Probability Deviation is always greater than or equal to 0.
If the deviation becomes 0, it means that the probabilities of the two answers are identical.

\textbf{PST} (Probability of Successful Trial) \cite{saravanan}: This metric refers to the ratio of successful measurements (readout of correct answers) to the total number of measurements.
For instance, the PST of 0.5 indicates that the half of the total measurement attempts resulted in one of the correct answer states for that quantum benchmarks.

\textbf{Hellinger Distance} \cite{hellinger1909neue}: This metric is used to quantify the similarity between the probability distributions of the quantum program results.
Hellinger Distance could be calculated as follows:

\begin{equation*} 
\centering
H(P, Q) = \sqrt{\cfrac{1}{2} \sum_{i=1}^{N} (\sqrt{p_i} - \sqrt{q_i})^2}
\end{equation*}

where $P = \{p_1, p_2, \ldots, p_N\}$ and $Q = \{q_1, q_2, \ldots, q_N\}$ are the probability distributions, and $p_i$ and $q_i$ are the $i$-th probabilities in $P$ and $Q$, respectively.
Hellinger Distance ranges from 0 to 1, where 0 indicates that the two probability distributions are identical.
In this work, we compare distributions $P$ obtained from quantum devices against probability distributions $Q$ obtained from using noise-less quantum circuit simulator (i.e., \textit{qasm\_simulator}) for each program.


\section{Results and Analysis}

\subsection{Deviation Analysis for 2-Answer Workloads}

\begin{table}[h]
\centering
\renewcommand{\arraystretch}{1}
\renewcommand{\tabcolsep}{0.6mm}
\caption{
Probability Deviations in 2-Answer Workloads}
\label{t3}
\begin{tabular}{|c|c|c|c|c|}
\hline
\multirow{2}{*}{Workload} & Standard & Bit-Inverted & Barber & Deviation Reduction   \\
 & Results & Results & Results & (over standard) \\
\hline
\hline
GHZ\_6 & 27\% & 32\% & 3\% &  89\% \\
GHZ\_9 & 78\% & 70\% & 4\% & 95\%  \\
GHZ\_12  & 81\% & 52\% & 19\% & 77\%  \\
\hline
GRV\_3b & 125\% & 95\% & 15\% & 88\%  \\
GRV\_4b & 80\% & 71\% & 5\% & 94\%  \\
\hline
MCR\_4 & 39\% & 37\% & 1\% & 97\%  \\
MCR\_5 & 73\% & 53\% & 13\% & 82\%  \\
MCR\_6  & 63\% & 51\% & 8\% &  87\% \\
\hline
MCS\_4 & 50\% & 32\% & 14\% & 72\%  \\
MCS\_5 & 47\% & 56\% & 6\% &  87\% \\
MCS\_6 & 83\% & 114\% & 17\% & 80\%  \\
\hline
Gmean & 63\% & 56\% & 7\% & 86\%  \\
\hline
\end{tabular}
\end{table}

\noindent
\cref{t3} shows the probability deviations between correct answer states from the standard, bit-inverted, Barber circuit execution results for 2-answer quantum circuit benchmarks (lower is better).
At \cref{t3}, the probability deviation between correct answer states is 63\% for the standard circuit execution and 56\% for the bit-inverted circuit execution on the geometric mean, respectively.
It is shown that these probability deviations can be as large as 125\% (for the case of GRV\_3b) in the worst-case standard circuit execution.
On the other hand, when Barber is applied, the probability deviation between correct answer states shows 7\% on the geometric mean (up to 19\%) and it is reduced by 86\% compared to the probability deviation of standard circuit execution.
These experimental results demonstrate that Barber can efficiently mitigate the probability deviation between correct answers in 2-Answer quantum programs.

\subsection{Comparative Analysis with Related Works}

\noindent
In this section, we evaluate the Probability of Successful Trials (PST \cite{saravanan}) and the Hellinger distance (relative to noise-free distributions  \cite{hellinger1909neue}) of quantum program result distributions achieved through various quantum error mitigation techniques. 
It is important to note that not all error mitigation techniques are applicable to every quantum circuit benchmark.
For instance, the Enhancing Virtual Distillation technique, which necessitates the parallel execution of two copies of the original quantum circuit \cite{li2023enhancing}, cannot be deployed on the \textit{ibm\_algiers} system if the qubit count of the workload exceeds 13 qubits.
Additionally, the Quixote cannot applied to GRV (Grover's algorithm circuit \cite{grover}) and QFT (Quantum Fourier Transform \cite{shor}) due to their circuit structures not meeting the sub-circuit partitioning requirements \cite{jang}. 
In such cases where these techniques cannot be applied, we reference data based on the results from standard circuit executions.
In contrast, the Barber could be applied to all kinds of the quantum circuit benchmarks.

\begin{figure} [h] 
\centerline {
\includegraphics [width=\columnwidth] {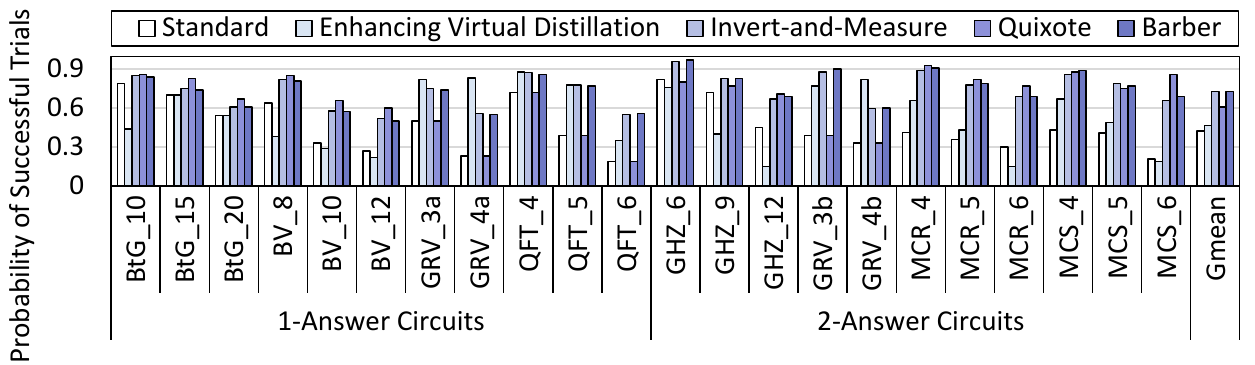} }
\caption {
The comparative evaluations of PST results for Barber's method with previous error mitigation techniques.
} 
\label{f2} 
\end{figure}

\textbf{PST Evaluations}:
\cref{f2} presents PST (Probability of Successful Trials) comparisons across various quantum error mitigation techniques.
Standard circuit execution yields an average PST of 0.42, indicating that when only the Qiskit Runtime Sampler is employed, the current quantum system (\textit{ibm\_algiers}) produces correct answers with a probability of less than half for the given benchmark workloads.
Enhancing Virtual Distillation offers a 10\% improvement in PST over standard execution; however, it exhibits PST degradation for some workloads over the 6-qubit scale.
This degradation can occur because duplicating and executing quantum circuits through Enhancing Virtual Distillation increases the program size, making the system more vulnerable to cross-talk errors between qubits.
Both Invert-and-Measure and Barber achieve an average PST improvement of 0.73 over standard execution, marking the highest gain among the tested techniques.
This improvement is attributed to reducing the probabilities of incorrect states when merging and normalizing the standard and inverted executions.
Quixote shows an average PST of 0.61 due to reducing program size through circuit partitioning (when only it is applicable), thereby mitigating errors.

\begin{figure} [h] 
\centerline {
\includegraphics [width=\columnwidth] {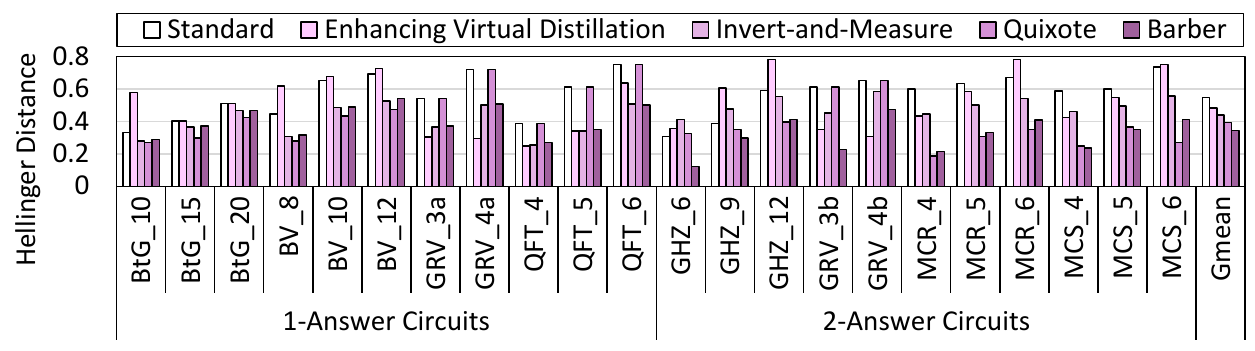} }
\caption {
The comparative evaluations of the Hellinger Distances for Barber with previous error mitigation techniques.
} 
\label{f3} 
\end{figure}

\textbf{Hellinger Distance Evaluations}:
\cref{f3} compares Hellinger distances for various quantum error mitigation techniques.
Barber achieves the lowest Hellinger distance of 0.34 across five execution scenarios, which is 15\% lower than Quixote, 28\% lower than Invert-and-Measure, 41\% lower than Enhancing Virtual Distillation, and 60\% lower than standard circuit execution.
Specifically, for 2-answer circuits, Barber achieves an average Hellinger distance of 0.29, which is 17\% lower than Quixote, 67\% lower than Invert-and-Measure, 72\% lower than Enhancing Virtual Distillation, and 90\% lower than standard circuit execution.
Quixote achieves the second-lowest average Hellinger distance due to its approach of reducing the depth and size of quantum programs through circuit partitioning, thus minimizing the impact of thermal relaxation.
On the other hand, Invert-and-Measure achieves a higher average PST than Quixote but a higher Hellinger distance. 
This discrepancy may be attributed to Invert-and-Measure augmenting the probability deviation between correct answer states during the reconstruction of the original results.
Barber shows an equivalent PST to Invert-and-Measure but achieves a lower Hellinger distance, indicating that Barber can offset the probability deviation between answers.

\textbf{Comprehensive Analysis of Result Distributions}:
As shown in \cref{f2,f3}, it is observed that the proportional relationship between PST and Hellinger distance varies depending on the number of correct answer states in the quantum workload.
For 1-answer circuits, PST and Hellinger distance are inversely proportional: higher PST values correspond to lower Hellinger distances.
This occurs because, in 1-answer circuits, measuring the single correct state with high probability increases the similarity straightly with the noise-free results.
However, for 2-answer circuits, a higher PST does not necessarily correlate with a lower Hellinger distance.
This is because, even if the correct states are measured with high probability, the similarity with noise-free results can be relatively low if the probability deviation between the correct answer states becomes large.
Barber shows the highest PST and the lowest Hellinger distance among the error mitigation techniques, demonstrating that Barber efficiently mitigates the probability deviation between answers while outputting the answer states with higher probabilities.

\section{Conclusion}

\noindent
Barber is a quantum compilation technique designed to mitigate distortions in the output probability distribution of near-term quantum programs caused by thermal relaxation errors of qubits. 
Barber achieves this by balancing probability deviations between correct answer states by combining results from the standard quantum circuit and its bit-inverted counterpart, which experiences thermal relaxation in the opposite direction. 
Barber is compatible with all types of gate-based quantum computers without requiring a hardware modification and could be applied to any quantum algorithm.

\begin{acks}
This research was funded by the National Research Foundation of Korea (NRF), supported by the Korean government (Ministry of Science and ICT (MSIT)) under the project Creation of the Quantum Information Science R\&D Ecosystem based on Human Resources (No. RS-2023-00303229). 
Won Woo Ro is the corresponding author, and correspondence regarding this work should be directed to him.
\end{acks}


\bibliographystyle{ACM-Reference-Format}
\bibliography{sample-base}

\end{document}